\documentclass[conference]{IEEEtran}
\ifCLASSINFOpdf
\else
\fi

\usepackage{array}
\usepackage{amsmath}
\usepackage{amssymb}
\usepackage{setspace}
\usepackage{graphicx}
\usepackage[tight,footnotesize]{subfigure}

\begin{document}

\title{VDAS: Verifiable Data Aggregation Scheme for Internet of Things}

\author{\IEEEauthorblockN{Jingwei Liu\IEEEauthorrefmark{1},
Jinping Han\IEEEauthorrefmark{1},
Longfei Wu\IEEEauthorrefmark{2},
Rong Sun\IEEEauthorrefmark{1} and
Xiaojiang Du\IEEEauthorrefmark{2}}
\IEEEauthorblockA{\IEEEauthorrefmark{1}State Key Lab of ISN, Xidian University, Xi'an, 710071, China.\\ Email: jwliu@mail.xidian.edu.cn, 18292003217@163.com, rsun@mail.xidian.edu.cn}
\IEEEauthorblockA{\IEEEauthorrefmark{2}Department of Computer and Information Sciences, Temple University, Philadelphia, PA 19121, USA.\\
Email: longfei.wu@temple.edu, dxj@ieee.org}
}
\maketitle

\begin{abstract}
Along with the miniaturization of various types of sensors, a mass of intelligent terminals are gaining stronger sensing capability, which raises a deeper perception and better prospect of Internet of Things (IoT). With big sensing data, IoT provides lots of convenient services for the monitoring and management of smart cities and people's daily lives. However, there are still many security challenges influencing the further development of IoT, one of which is how to quickly verify the big data obtained from IoT terminals. Aggregate signature is an efficient approach to perform big data authentication. It can effectively reduce the computation and communication overheads. In this paper, utilizing these features, we construct a verifiable data aggregation scheme for Internet of Things, named VDAS, based on an improved certificateless aggregate signature algorithm. In VDAS, the length of the aggregated authentication message is independent of the number of IoT terminals. Then, we prove that VDAS is existentially unforgeable under adaptive chosen message attacks assuming that the computational Diffie-Hellman problem is hard. Additionally, the proposed VDAS achieves a better trade-off on the computation overheads between the resource-constrained IoT terminals and the data center.
\end{abstract}
\IEEEpeerreviewmaketitle

\section{Introduction}
Today, as the traditional Internet is merging with different kinds of wired/wireless networks, Internet of Things (IoT) has attracted the attention worldwide by enabling things-to-things and things-to-people communications. IoT can be widely used in ubiquitous applications of smart cities, including smart building, smart grid, public transportation, health-care, etc. With the latest advances on a diversity of sensors integrated in the intelligent terminals, IoT has opened new opportunities for the development of various industries to improve people's daily lives.

\begin{figure}[tb]
  \centering
  \includegraphics[width=8.5cm]{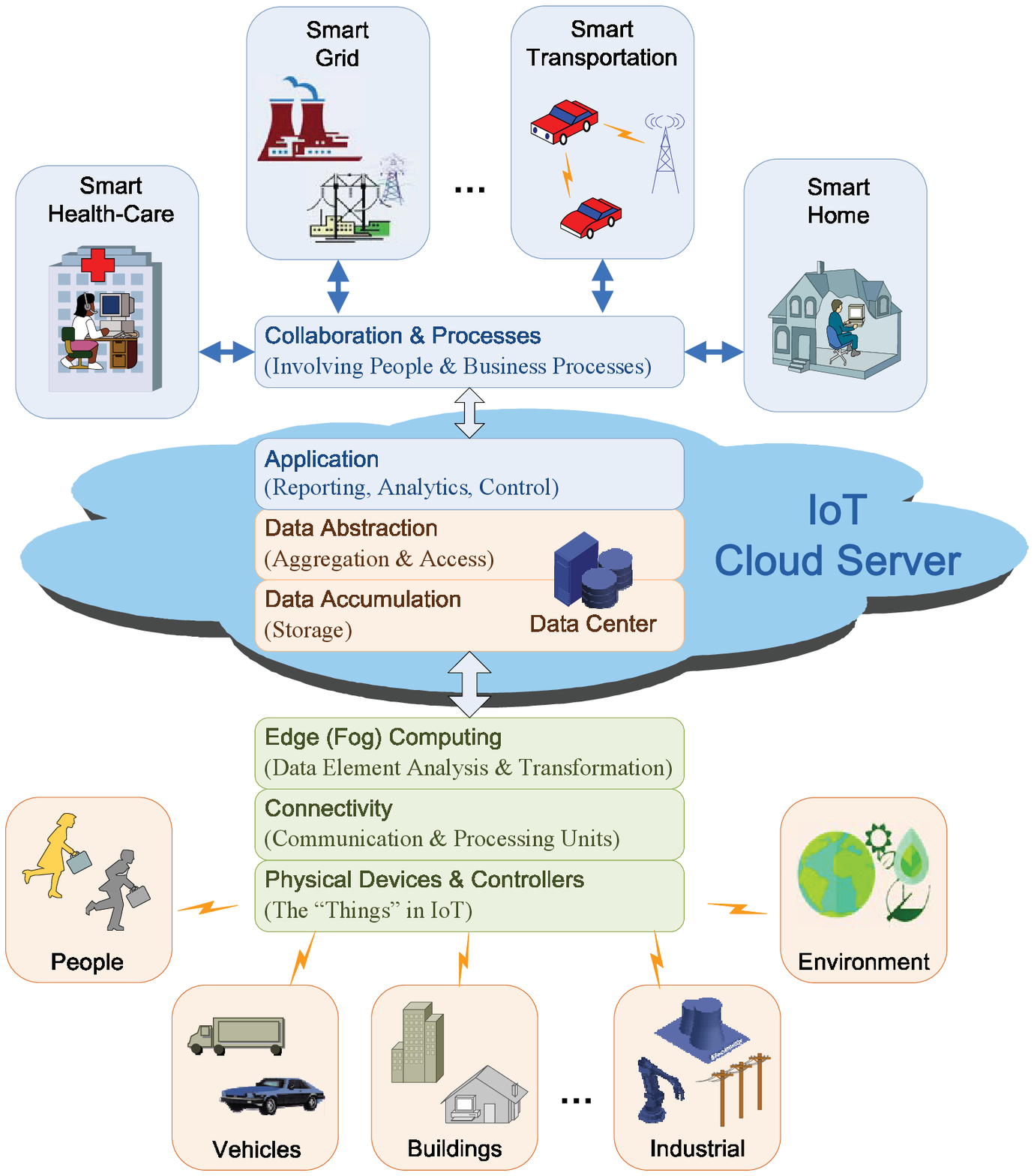}\\
  \caption{Seven-level reference model}
  \label{model}
\end{figure}

We have seen a vast amount of works on every aspect in the IoT systems \cite{TAFGNOPG16, NJ16, YHDZ13, GBMP13, AIM10, CISCO14, DGXC08, HDWH10}, including the reliability, flexibility, robustness, and security of IoT, to make environmental conditions more controllable, convenient, and safe. To better understand IoT, three reference models have been wildly discussed: three-level model \cite{GBMP13}, five-level model \cite{AIM10}, and seven-level model \cite{CISCO14}. The seven-level model was proposed by CISCO in 2014, as shown in Fig. \ref{model}, and is broadly accepted by industry and academia.

Nowadays, IoT has been playing an important role of bridging the physical world and the digital world. By integrating all relevant technologies together (such as computer networks, wireless sensor networks, cellular networks, etc.), IoT allows people and things to be connected with each other anytime, anywhere. For the intrinsic features of IoT-based applications, IoT has raised lots of new security challenges \cite{TAFGNOPG16}, including: a. correctly sensing the environment; b. securely exchanging the information; c. safeguarding the private information. Additionally, based on CISCO's seven-level reference model, Nia and Jha summarized the security threats and corresponding countermeasures in IoT, level by level \cite{NJ16}. Lots of efforts were focused only on the lightweight security solution in the three lower levels \cite{HD11, YHD14, HDLLS15}. Moreover, a massive amount of IoT applications in smart cities neglect to verify the obtained data, especially when huge amount of data are generated by a large scale of devices. The protection and verification of the huge of IoT data must be taken into consideration \cite{ZZZD15, HD14}. Aggregate Signature (AS) is an efficient approach to batch verification of massive data in IoT and raises a better prospect of IoT.

The concept of aggregate signature was first introduced by Boneh et al. \cite{BGL03} in Eurocrypt 2003. This kind of digital signature can aggregate \emph{n} signatures on \emph{n} distinct messages from \emph{n} individual users into a single signature, which allows the aggregator to easily verify that the \emph{n} users have indeed signed the \emph{n} original messages. Since the scheme can greatly reduce the total signature length and the verification overhead, it is very useful especially in the environments with low communication bandwidth, low storage and low computing power. The typical applications include wireless sensor networks, vehicular communications and some other IoT scenarios. Taking advantages of these merits, many AS schemes based on the Traditional Public Key Cryptosystem (TPKC), ID-based Public Key Cryptosystem (ID-PKC) and Certificateless Public Key Cryptosystem (CL-PKC) have been proposed for various applications in practical circumstances.

In 2004, Cheon et al. \cite{CKY04} introduced the first identity-based aggregate signature. Soon after, many certificateless aggregate signature (CL-AS) schemes \cite{LLS14, XGCL13, ZQW10, CWZ15, CSZ10, DHW13, CTM15, KX16, GLHC07} were proposed respectively, due to CL-PKC's resistance to the key escrow problem in ID-PKC. In \cite{LLS14}, the authors proposed a CL-AS scheme that requires only four pairing computations for aggregate verification and two group elements in signature size, but it fails to provide unforgeability. In \cite{XGCL13}, the authors proposed an efficient CL-AS scheme with better performance than the previous schemes \cite{ZQW10, GLHC07}. Unfortunately, this scheme was proved to be vulnerable to the attack launched by a Type II adversary in \cite{HTC12}.

In this paper, based on an improved CL-AS algorithm, we propose an efficient verifiable data aggregation scheme for IoT, named VDAS. It effectively reduces the computation overhead in IoT data center. Additionally, the proposed VDAS is proved to be secure assuming that the computational Diffie-Hellman (CDH) problem is hard. The length of the aggregated authentication message in VDAS is only two group elements, and is independent of the number of signers. Moreover, VDAS achieves a lower computation overhead in the individual signing phase and aggregate verification phase, which is more compatible and preferred by the resource-limited IoT terminal devices and the data center.

The rest of this paper is organized as follows. In section II, we briefly introduce some preliminaries including pairing, the computational assumption, and the security model. In section III, we first describe VDAS in detail, and then formally analyze its security. In section IV, the performance is evaluated. Finally, the conclusion is given in section V.

\section{Preliminaries}
To facilitate the understanding of the cryptogram essential, we introduce the basic definitions and the properties of bilinear pairings over elliptic curve group. We also give the security model for VDAS.

\subsection{Bilinear Pairings}
\noindent \textbf {Definition 1.} \emph{Bilinear Pairings map}: $G_{1}$ and $G_{2}$ are the cycle additive group and cycle multiplicative group of prime order $q $ respectively. $P$ is a generator of $G_{1}$. A bilinear pairing is a map $e \colon G_{1}\times G_{1}\to G_{2}$ , it satisfies the following properties:

\begin{itemize}
\item Bilinearity: For any $P$, $Q \in G_{1}$, random number $a$, $b\in {\rm Z}_q^\ast$, we have $e(aP, bQ) = e(P, Q)^{ab}$;

\item Non-degeneracy: There exists $P$, $Q \in G_{1}$, such that $e(P, Q) \neq 1$;

\item Computability: There exists an efficient polynomial time algorithm to compute $e(P, Q)$, for any $P$, $Q \in G_{1}$;
\end{itemize}

\noindent \textbf{Definition 2.} \emph{ computational Diffie-Hellman (CDH) Problem}: $G_{1}$ is a cycle additive group of prime order $q$, $P$ is the generator of $G_{1}$, for any $a$, $b\in {\rm Z}_q^\ast$, given an instance $\langle P,aP,bP \rangle$, compute ${abP}\in G_{1}$.

\subsection{Security Model}
In general, a CL-AS scheme contains six algorithms: \emph{Setup}, \emph{Partial-Private-Key-Extract}, \emph{UserKeyGen}, \emph{Sign}, \emph{Aggregate} and \emph{Aggregate Verify}. There are two types of attackers in CL-AS: $\mathcal{A}_{I}$ and $\mathcal{A}_{II}$. $\mathcal{A}_{I}$ is able to replace any user's public key, while $\mathcal{A}_{II}$ could be an honest-but-curious Key Generation Center (KGC) who holds the master-key but is unable to perform public key replacement attack. To prove the security of VDAS based on CL-AS, we adopt the existential unforgeability under the adaptive-chosen-message attacks and adaptive-chosen-identity attacks model in \cite{HMS11} for both types of adversaries. There is no probabilistic polynomial time adversary, no matter $\mathcal{A}_{I}$ or $\mathcal{A}_{II}$, could win the game with non-negligible probability.

\begin{figure}[tb]
\begin{center}
\includegraphics[width=8.5cm]{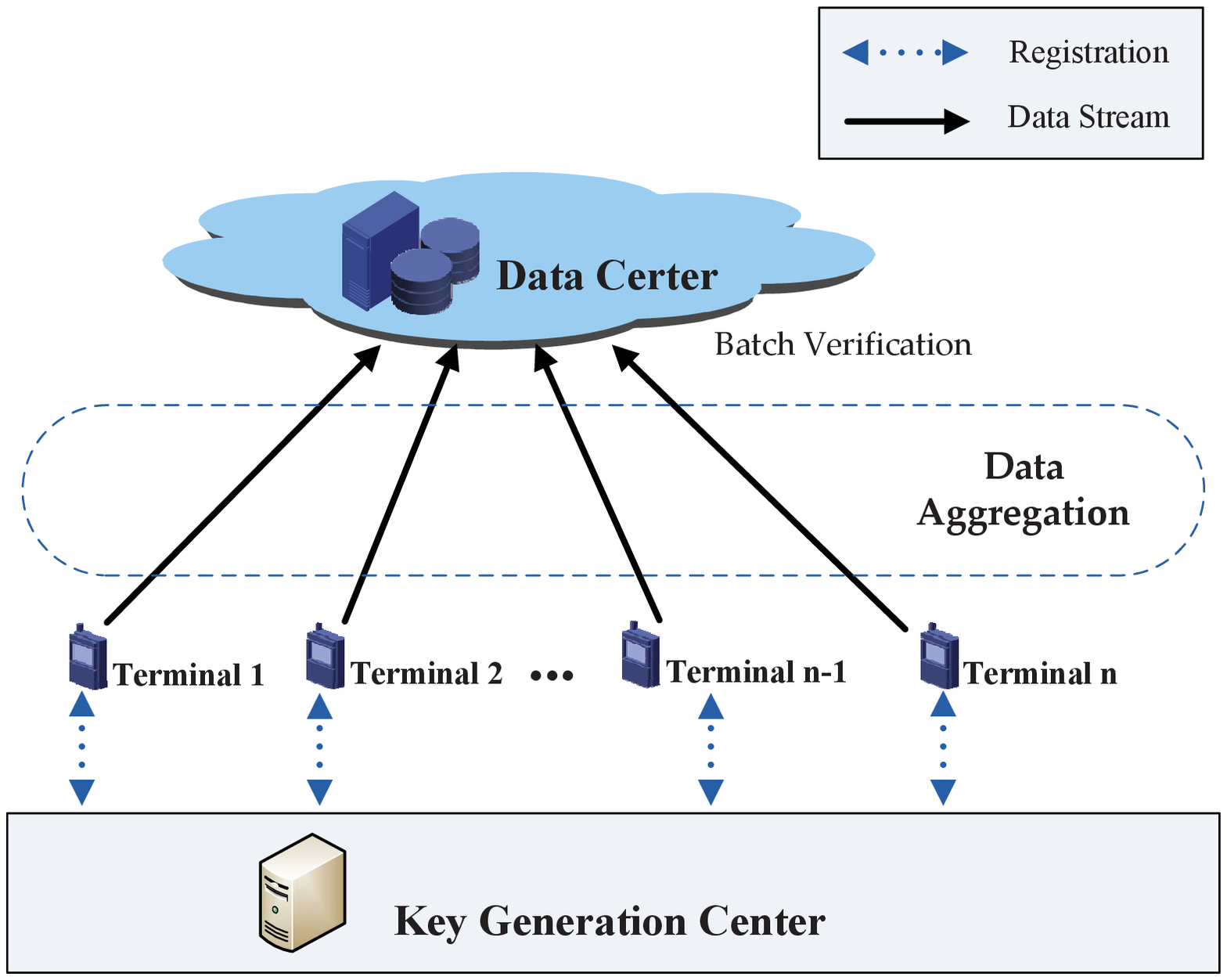}\\
\caption{The network architecture of VDAS}\label{VDAS}
\end{center}
\end{figure}

\section{An Efficient Verifiable Data Aggregation Scheme for IoT}
As IoT is becoming more autonomous in smart cities, a key challenge for IoT towards smart city applications is how to make sure a large amount of data indeed come from the legitimate devices. Once the IoT data center gathers huge amount of data in real time, it may not be able to efficiently verify the collected data.

\subsection{Design Objectives}
On account of the heavy burdens for the verification of a mass of gathered data in IoT scenarios, we propose an efficient verifiable data aggregation scheme that can reduce the computation overhead for the IoT data center. Deploying an improved CL-AS as the cryptogram essential, our scheme provides batch authentication of the collected data. In VDAS, there are three types of entities involved, as shown in Fig. \ref{VDAS}: $n$ IoT terminals, a Key Generation Center (KGC), and a IoT data center (as the aggregator). In general, we assume that all three kinds of entities can run relevant public key cryptographic operations. This implies the existence of some authority mechanisms, such as the KGC can generate and certify the cryptographic keys. The IoT terminal devices, as well as the data center, must contact the KGC in advance for key distribution.

\subsection{Verifiable Data Aggregation Scheme for IoT}
In this section, we will describe the proposed VDAS in detail, which consists of four stages: \emph{System Setup}, \emph{Registration}, \emph{Individual Signing}, and \emph{Aggregate Verification}. To enhance the security of the scheme, we introduce a state information $\Delta$ as defined in \cite{ZQW10, CWZ15}, which is a stochastic-length bit string selected randomly by terminals and broadcasted before the signing phase. One can choose the current time, some parts of the system parameters or other feasible information to generate the $\Delta$. The specification of the scheme is as follows:

\begin{itemize}
\item[1)]{\emph{System Setup}}: Given a security parameter $l$, the KGC chooses a cyclic additive group $(G_{1}, +)$ which is generated by $P$ with prime order $q (q > 2^{l})$, a cyclic multiplicative group $(G_{2}, \cdot)$ of the same order, and a bilinear map $e: G_{1} \times G_{1} \to G_{2}$. The KGC also selects two cryptographic hash functions $\textsf{H}_{1}: \{0,1\}^* \to G_{1}$, $\textsf{h}_{2}: \{0, 1\}^* \to {\rm Z}_q^*$. Then, it picks a random $s \in {\rm Z}_q^\ast$ as the master-key and accordingly sets $P_{0} = sP$. The system parameters are $Param = \{G_{1}, G_{2}, e, q, P, P_{0}, \textsf{H}_{1}, \textsf{h}_{2}\}$ and the data space is $data \in \{0,1\}^*$. Eventually, the KGC publishes $Param$ while keeping the master-key in secret.

\item[2)]{\emph{Registration}}: Upon receiving a registration request from a terminal $i$, KGC first confirms the identity $ID_{i}\in\{0,1\}^*$. Then, it computes $Q_{i} = \textsf{H}_{1}(ID_{i})$ and generates the partial private key $D_{i} = sQ_{i}$ for the terminal. The terminal also generates a random $x_{i} \in {\rm Z}_q^\ast$ as its secret value, computes $P_{i} = x_{i}P$, and sets $P_{i}$ as its public key, $\langle x_{i}, D_{i} \rangle$ as its private key. Finally, KGC sends $\langle ID_i, Q_i, P_i \rangle$ to the data center.

\item[3)]{\emph{Individual Signing}}: Based on the common state information $\Delta$, the terminal $i$, whose identity is $ID_{i}$ and the corresponding public key is $P_{i}$, signs a requested $data_{i}$ with its private key $\langle x_{i}, D_{i} \rangle$ as follows:
    \begin{itemize}
        \item Choose a random $r_{i} \in {\rm Z}_q^\ast$, then compute $R_{i} = r_{i}P$, $h_{i} = \textsf{h}_{2}(data_{i} \parallel \Delta \parallel ID_{i})$, and $g_{i} = \textsf{h}_{2}(data_{i} \parallel \Delta \parallel P_{i})$;
        \item Compute $V_{i} = g_{i}D_{i} + (x_{i}h_{i} + r_{i})U$, in which $U = H_{1}(\Delta \parallel P_{0})$;
        \item Upload $\sigma_{i} = (R_{i}, V_{i})$ as the signature on $data_{i}$ to the data center.
    \end{itemize}

\item[4)]{\emph{Aggregate Verification}}: Upon receiving a large number of $data$ needed to be verified, the aggregator, that is the data center, aggregates a collection of individual signatures under the same state information $\Delta$. For $n$ terminals with identities $L_{ID}=\{ID_{1}, ID_{2},\ldots,ID_{n}\}$, the corresponding public keys are $L_{PK} = \{P_{1}, P_{2}, \ldots, P_{n}\}$ and data-signature pairs are $\langle data_{1}, \sigma_{1} \rangle, \langle data_{2}, \sigma_{2} \rangle, \ldots, \langle data_{n}, \sigma_{n} \rangle$, respectively. The IoT data center computes $R = R_{1} + R_{2} + \ldots + R_{n}$, $V = V_{1} + V_{2} + \ldots + V_{n}$ and outputs $\langle R, V \rangle$ as the aggregated signature $\sigma$. To verify if the aggregate signature $\sigma = (R, V)$ on all messages $\langle data_{1}, data_{2}, \ldots, data_{n} \rangle$ is signed by $n$ terminals with the same state information $\Delta$, the data center performs the following steps:
\begin{itemize}
    \item Compute $U = \textsf{H}_{1}(\Delta \parallel P_{0})$;
    \item For all $1\leqslant i\leqslant n$, compute $h_{i} = \textsf{h}_{2}(data_{i} \parallel \Delta \parallel ID_{i})$, $g_{i} = \textsf{h}_{2}(data_{i} \parallel \Delta \parallel P_{i})$;
    \item Verify $e(V, P) = e(\sum_{i = 1}^{n} g_{i}Q_{i}, P_{0}) e(\sum_{i = 1}^{n} h_{i}P_{i} + R, U)$. If the equation holds, output true. Otherwise, output false.
 \end{itemize}
\end{itemize}

\begin{figure}[tb]
\begin{center}
\includegraphics[width=8cm]{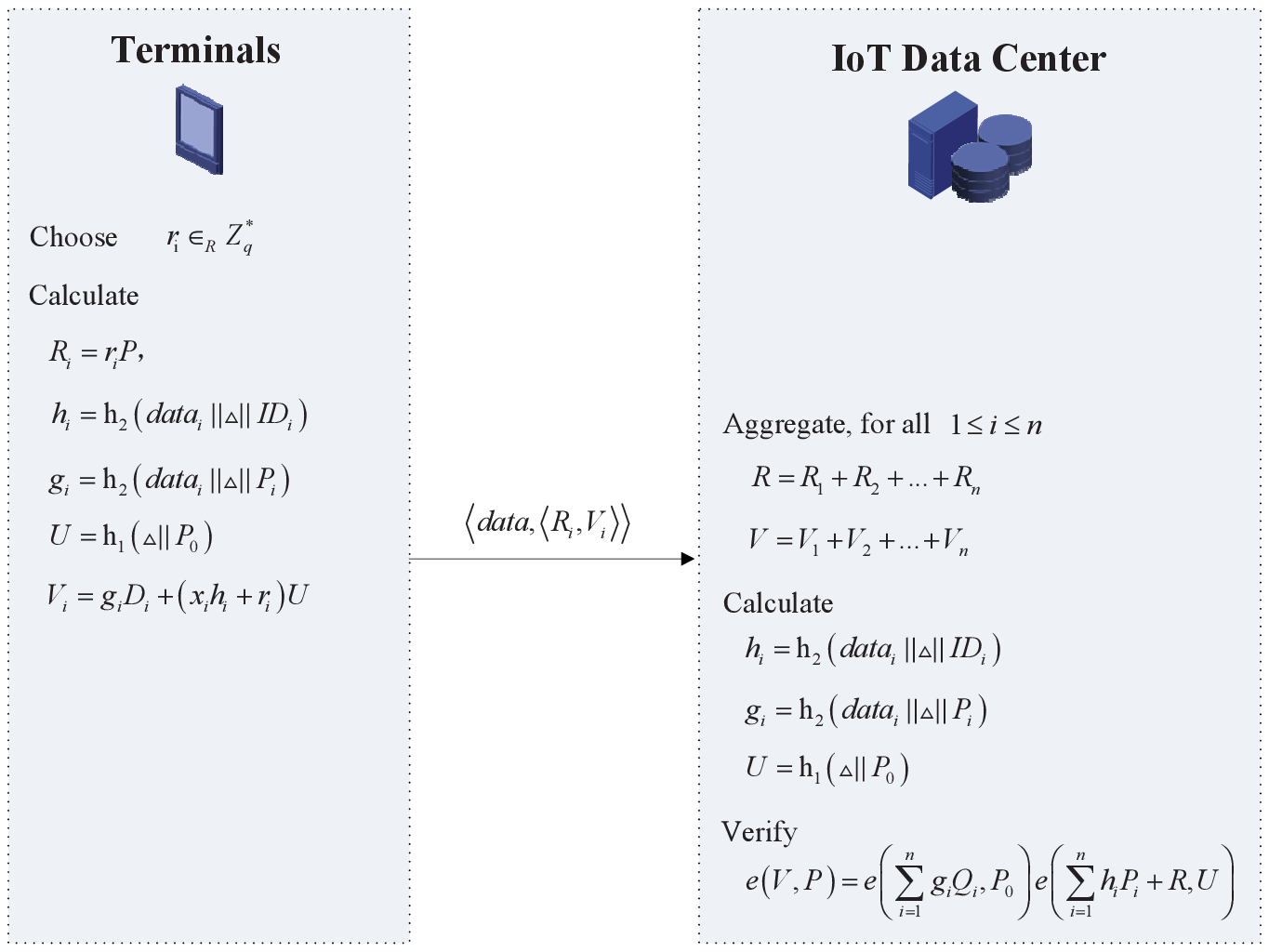}\\
\caption{Verifiable data aggregation}\label{Aggregate}
\end{center}
\end{figure}

\subsection{Security Analysis}

\noindent \textbf{Theorem 1.} \emph {The proposed VDAS is existential unforgeable against the Type I adversaries $\mathcal{A}_I$ in the random oracle model assuming the CDHP is hard.}

\noindent \textbf{Proof:} {Let $\mathcal{C}$ be a challenger who receives a random instance $(P,aP,bP)$ of the CDH problem in $(G_{1},+)$. Adversary $\mathcal{A}_{I}$ is allowed to ask all oracles maintained by $\mathcal{C}$. We will show how $\mathcal{C}$ interact with $\mathcal{A}_{I}$ to compute $abP$ to solve the CDH problem with a non-negligible probability.}

\noindent \emph{\textbf{Setup}}: Firstly, $\mathcal{C}$ sets $P_{0}=aP$ and sends $Param = \{G_{1}, G_{2}, e, q, P, P_{0}, \textsf{H}_{1}, \textsf{h}_{2}\}$ to $\mathcal{A}_{I}$.

\noindent \emph{\textbf{Simulation}}: $\mathcal{C}$ operates the hash functions $\textsf{H}_{1}$ and $\textsf{h}_{2}$ as two random oracles. $\mathcal{A}_{I}$ can perform the following different types of queries in an adaptive manner. $\mathcal{C}$ needs to maintain several lists generated by $\textsf{H}_{1}$ and $\textsf{h}_{2}$, which are initially empty.

\emph{Register-query}: Here, $\mathcal{C}$ selects a $j \in \{1, 2,\ldots, n\}$ and $\mathcal{A}_{I}$ can make \emph{Register-query} on any $ID_i$. $\mathcal{C}$ selects $x_{i}, y_{i} \in _{R}{\rm Z}_q^\ast$ randomly, where $y_{i}$ is not in the list $L_R$. If $i \neq j$, $\mathcal{C}$ computes $Q_{i} = y_{i}P$, $P_{i} = x_{i}P$, $D_{i} = y_{i}P_{0}$; Otherwise, $\mathcal{C}$ randomly chooses a $Q_{j} = bP \in G_1$, and set $D_{j} = "unknown"$, $P_{j} = x_{j}P$. Finally, $\mathcal{C}$ adds the relevant information to the list $L_R$ and returns $Q_{i}$ and $P_{i}$ to $\mathcal{A}_{I}$.

\emph{Partial-Private-Key-query}: $\mathcal{A}_{I}$ can ask questions on any $ID_i$. If $i \neq j$, $\mathcal{C}$ searches the list $L_R$ for the relevant information on $ID_{i}$, and returns $D_{i}$ to $\mathcal{A}_{I}$. Otherwise, $\mathcal{C}$ aborts.

\emph{Public-Key-Replace-query}: $\mathcal{A}_{I}$ can make this query on $\langle ID_{i},P_{i}^\prime \rangle$, $\mathcal{C}$ checks the list $L_R$, replacing $\langle P_{i}, x_{i}, y_{i} \rangle$ with $\langle P_{i}^\prime, ``unknown", y_{i} \rangle$.

\emph{Secret-Value-query}: $\mathcal{A}_{I}$ can query on any $ID_i$. If the corresponding $x_{i} \neq ``unknown"$ is in the list $L_R$, $\mathcal{C}$ returns $x_{i}$ to $\mathcal{A}_{I}$; Otherwise, it outputs $``unknown"$.

$\textsf{H}_{1}$-\emph{query}: $\mathcal{A}_{I}$ can query on $\langle \Delta_i, P_{0} \rangle$. $\mathcal{C}$ selects $\lambda_{i} \in _{R}{\rm Z}_q^\ast$, which is not in the list $L_{\textsf{H}_{1}}$. Then, it sets $U_{i} = \lambda_{i}P - P_{0}$, and sends it to $\mathcal{A}_{I}$. Finally, $\langle \Delta_i, U_{i}, \lambda_{i} \rangle$ will be added to the list $L_{\textsf{H}_{1}}$.

$\textsf{h}_{2}$-\emph{query}: $\mathcal{A}_{I}$ can make this query on any $(0, 1)^*$, $\mathcal{C}$ selects $h_{i}$ or $g_{i} \in _{R}{\rm Z}_q^\ast$, which is in the list $L_{\textsf{h}_{2}}$. Then, it returns $h_{i}$ or $g_{i}$ to $\mathcal{A}_{I}$ and adds the relevant information to the list $L_{\textsf{h}_{2}}$.

\emph{Individual-Signing-query}: when $\mathcal{A}_{I}$ asks for signing on any $\langle data_{i}, \Delta_i, ID_{i}, P_{i} \rangle$, $\mathcal{C}$ performs as follows:
\begin{enumerate}
  \item [(1)] Select $r_{i},h_{i},g_{i}\in _{R}{\rm Z}_q^\ast$ randomly, while $h_{i}$ and $g_{i}$ are not in the list $L_{\textsf{h}_{2}}$.

  \item [(2)] Compute $R_{i}=r_{i}P+g_{i}Q_{i}-h_{i}P_{i}$.

  \item [(3)] Search $U_{i} = \lambda_{i}P - P_{0}$ in $L_{\textsf{H}_{1}}$ and compute $V_{i} = \lambda_{i}g_{i}Q_{i} + r_{i}\lambda_{i}P - r_{i}P_{0}$.
\end{enumerate}

$\mathcal{C}$ adds all above information to the corresponding list and returns $\langle R_{i}, V_{i} \rangle$ to $\mathcal{A}_{I}$. $\langle R_{i}$, $V_{i}\rangle$ could be easily proven to be valid, since
\begin{eqnarray}
\nonumber & & e(g_{i}Q_{i},P_{0})e(h_{i}P_{i}+R_{i},U_i)\\
\nonumber &=& e(g_{i}Q_{i},P_{0})e(r_{i}P+g_{i}Q_{i},\lambda_{i}P - P_{0})\\
\nonumber &=& e(\lambda_{i}r_{i}P+\lambda_{i}g_{i}Q_{i},P)e(r_{i}P, -P_{0})\\
\nonumber &=& e(\lambda_{i}g_{i}Q_{i} + r_{i}\lambda_{i}P-r_{i}P_{0}, P)\\
\nonumber &=& e(V_{i},P)
\end{eqnarray}
\noindent \emph{\textbf{Forgery}}: Finally, $\mathcal{A}_{I}$ returns a forged aggregate signature $\sigma=(R,V)$ on messages $\langle data_{1}, data_{2}, \ldots, data_{n} \rangle$, which is generated by $n$ terminals whose identities are $L_{ID} = \{ID_{1}, ID_{2}, \ldots, ID_{n}\}$ and corresponding public keys are $L_{PK} = \{P_{1}, P_{2}, \ldots, P_{n}\}$ with the same state information $\Delta$. Furthermore, the aggregate signature $\sigma$ satisfies the following conditions:
\begin{enumerate}
  \item [(1)] $e(V,P)=e(\sum_{i = 1}^{n}g_{i} Q_{i},P_{0})e(\sum_{i = 1}^{n}h_{i} P_{i} + R, U)$.
  \item [(2)] There is at least an identify $ID_{k} \in L_{ID}$, which has neither made \emph{Partial-Private-Key-query} nor \emph{Individual-Signing-query} on $\langle data_{j}, \Delta, ID_{j}, P_{j} \rangle$.
\end{enumerate}
From the Forking lemma \cite{DJ00}, if $\mathcal{C}$ has a replay with the same random tape but a different respond of $\textsf{h}_{2}$, $\mathcal{A}_{I}$ will output a new effective forged signature $\sigma^\prime = \langle R, V^\prime \rangle$. In this process, $k\in\{1,2,\ldots,n\}$. If $i \in \{1, 2, \ldots, n\}\backslash\{k\}$, we always have $g_{i} = g_{i}^\prime$; otherwise, if $i = k$, we have $g_{k} \neq g_{k}^\prime$. Hence, the following two equations hold:
\begin{equation}
 \begin{split}\nonumber
 \begin{cases}
    e(V, P) = e(\sum_{i = 1}^{n}g_{i} Q_{i},P_{0})e(\sum_{i = 1}^{n}h_{i} P_{i}+R, U)\\
    e(V^\prime, P) = e(\sum_{i=1}^{n}g_{i}^\prime Q_{i},P_{0})e(\sum_{i = 1}^{n}h_{i} P_{i}+R, U)
 \end{cases}
 \end{split}
\end{equation}
If $ID_{k} = ID_{j}$, then $Q_{k} = Q_{j} = bP$. $\mathcal{C}$ can output $abP$ as a solution to the CDH instance by computing $abP = (g_{s} - g_{s}^\prime)^{-1}(V_{s} - V_{s}^\prime)$ according to the above two equations. Otherwise, $\mathcal{C}$ aborts.

\noindent \textbf{Theorem 2.} \emph {The proposed VDAS is existential unforgeable against the Type II adversaries $\mathcal{A}_{II}$ in the random oracle model assuming the CDHP is hard.}

\noindent \textbf{Proof:} This security property also relies on the hardness of CDHP. It can be deduced similarly as the security proof of Theorem 1. Due to the page limitation, we omit the proof in detail.
\begin{table}
  \centering
  \caption{Time consumption on different basic cryptographic operations} \label{cryptographic operations}
  \tabcolsep 0.05in
  \setlength{\extrarowheight}{0.25cm}
  \footnotesize
  \begin{tabular}{c|ccc}
     \hline
     \quad \raisebox{0.1cm}{Operations} \quad & \quad \raisebox{0.1cm}{Multiplication} \quad & \quad \raisebox{0.1cm}{Hash} \quad & \quad \raisebox{0.1cm}{Pairing} \quad     \\
     \hline
     \quad \raisebox{0.1cm}{Time(ms)}   \quad & \quad \raisebox{0.1cm}{3.629}          \quad & \quad \raisebox{0.1cm}{0.477} \quad & \quad \raisebox{0.1cm}{4.359}  \quad     \\
     \hline
  \end{tabular}
\end{table}

\begin{table}
  \centering
  \caption{Time consumption on individual signing} \label{signing}
  \footnotesize
  \tabcolsep 0.05in
  \setlength{\extrarowheight}{0.25cm}
 \begin{tabular}{c|c c}
    \hline
    \quad \quad \raisebox{0.1cm}{Schemes} \quad \quad  & \quad \quad  \raisebox{0.1cm}{Time consumption (ms)} \quad \quad \\
    \hline
    ZQWZ\cite{ZQW10}  &   19.685\\
    CWZY\cite{CWZ15}  &   15.576\\
    CSZ\cite{CSZ10}   &   10.983\\
    DHW\cite{DHW13}   &   15.597\\
    CTMHH\cite{CTM15} &   15.576\\
    VDAS              &   11.371\\
    \hline
 \end{tabular}
\end{table}

\begin{table*}
  \centering
  \caption{Complexity comparison between different schemes} \label{Complexity}
  \setlength{\extrarowheight}{0.25cm}
  \footnotesize
  \tabcolsep 0.05in
  \begin{tabular}{c|ccc}
  \hline
  \quad \quad \quad \raisebox{0.1cm}{Schemes} \quad \quad \quad & \quad \quad \quad \raisebox{0.1cm}{Signing} \quad \quad \quad & \quad \quad \quad \raisebox{0.1cm}{Aggregate verification} \quad \quad \quad & \quad \quad \quad \raisebox{0.1cm}{Size} \quad \quad \quad \\
  \hline
  ZQWZ\cite{ZQW10}   &   $5S+3H$  &   $5P+2nS+(2n+3)H$        &   $2L$     \\
  CWZY\cite{CWZ15}   &   $4S+2H$  &   $4P+2nS+(n+2)H$         &   $(n+1)L$ \\
  CSZ\cite{CSZ10}    &   $3S$     &   $(n+1)P+2nS+nH$         &   $(n+1)L$ \\
  DHW\cite{DHW13}    &   $4S+2H$  &   $4P+2nS+(n+2)H$         &   $2L$     \\
  CTMHH\cite{CTM15}  &   $4S+2H$  &   $4P+2nS+2nH$            &   $2L$     \\
  VDAS               &   $3S+1H$  &   $3P+2nS+(n+1)H$         &   $2L$     \\
  \hline
  \end{tabular}
\end{table*}

\section{Performance Evaluation}
In this section, we will evaluate the performance of different schemes from a computational point of view and then we closely analyze the computation comparison between our scheme and the existing schemes \cite{ZQW10, CWZ15, CSZ10, DHW13, CTM15}. For the experimental evaluation, we set up a simulation environment to measure the computing time of the aforementioned schemes, particularly the computing time for the \emph{Individual Signing} phase and the \emph{Aggregate verification} phase. The details are as follows:

\noindent \textbf{(1) Environment setup}: The simulation environment is set up in Ubuntu 12.04 with an Intel(R) Pentium G630 2.70GHz processor and 4096MB memory. Each scheme will be run 100 times in order to compensate for the randomness of the results.

\noindent \textbf{(2) Simulation}: In the simulation procedure, the computation overhead is primarily caused by several kinds of cryptographic operations. In order to provide a brief estimation for the subsequent performance assessment, we mainly focus on the computation overhead of primary cryptographic processing. Firstly, we list the running time of several fundamental cryptographic operations in Table \ref{cryptographic operations}, such as the scalar multiplication in $G_{1}$, the pairing operation and the hash operation, which occupy a major computation overhead in all six selected schemes. Table \ref{signing} shows the time consumption in individual signing phase. Table \ref{Complexity} indicates the complexity comparison among different schemes. Here, ``$P$" denotes a pairing operation, ``$S$'' denotes a scalar multiplication in $G_{1}$, ``$L$" denotes the size of the elements in $G_{1}$, and ``$H$" denotes a hash operation $\{0,1\}^* \to G_{1}$. In the individual signing phase, the proposed VDAS involves only three scalar multiplications and one hash operation, while in the aggregate verification phase, it requires an increasing computation overhead with the number of IoT terminals. Table \ref{Complexity} shows that, except of the schemes in \cite{CWZ15, CSZ10}, the other four schemes have the fixed length of aggregated authentication message---$2L$. In the following part, more detailed results will indicate the trend of the total computation overheads with different user scale in the selected schemes.

\begin{figure}[tb]
\begin{center}
\includegraphics[width=9cm]{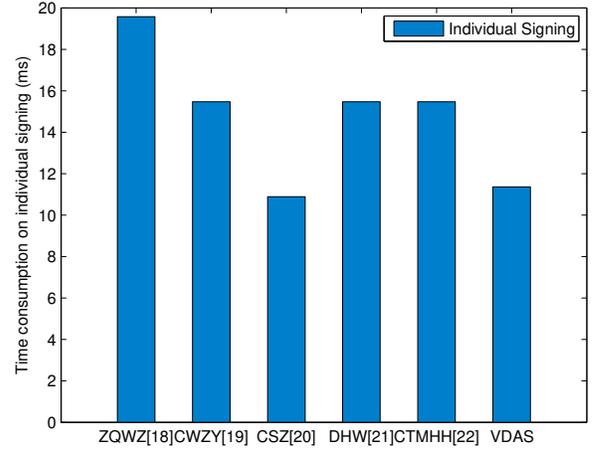}\\
\caption{Time consumption on individual signing}\label{individual_signing}
\end{center}
\end{figure}

\begin{figure*}[t]
\centerline{
\subfigure[Time consumption on signing vs. the number of terminals]{\includegraphics[width=8cm]{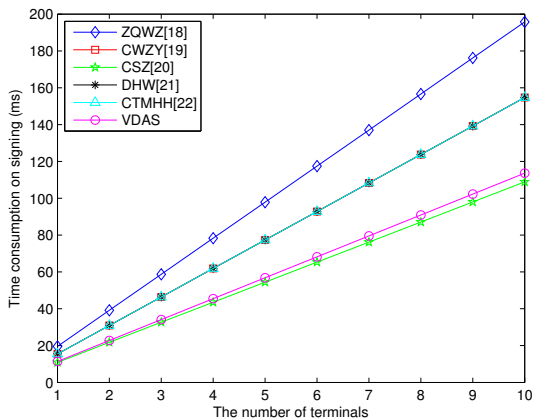}%
\label{signing_with_users}} %
%\hfil
\hspace{3mm}
%%%}
%%%\centerline{
\subfigure[Time consumption on aggregate verification vs. the number of terminals]{\includegraphics[width=8.2cm]{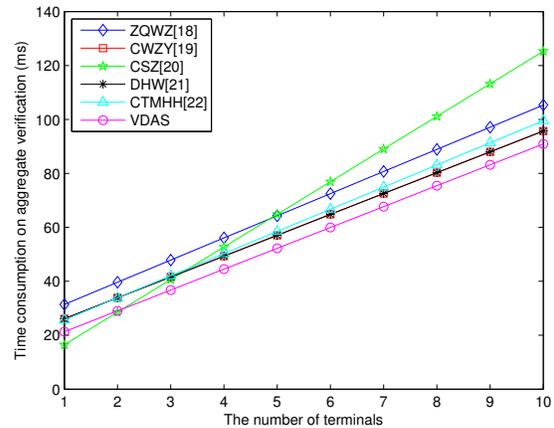}%
\label{verification_with_users}} %
}
% \captionsetup{skip=-1pt}
%%%\setlength{\abovecaptionskip}{-0.5 mm}
\setlength{\belowcaptionskip}{-2 mm}
\caption{Performance comparison between different schemes}
\label{comparison}
\vspace{-1mm}
\end{figure*}

Given the brief cryptographic operations and their corresponding time consumption, we can conveniently calculate the computation time in the ``individual signing'' phase and the ``aggregate verification'' phase respectively. Fig. \ref{individual_signing} shows that CSZ\cite{CSZ10} and VDAS achieve the better performance than the other selected schemes \cite{ZQW10, CWZ15, DHW13, CTM15} in the individual signing phase, and VDAS just needs a little bit more time than CSZ\cite{CSZ10}. However, CSZ\cite{CSZ10} is very sensitive to the number of IoT terminals in the aggregate verification phase. In spite of the good performance when $n \leq 5$, its computation overhead will sharply increase with the number of terminals, as shown in Fig. \ref{verification_with_users}. When $n > 5$, CSZ\cite{CSZ10} will spend much more time than the other schemes. In Fig. \ref{verification_with_users}, we also find that VDAS requires relatively less computation overhead than the other five selected schemes on aggregate verification. In IoT scenarios, the terminal devices are often energy-limited and computation-constrained while the IoT data center has heavy computation burden. Therefore, from the simulation results, the proposed VDAS achieves a more reasonable trade-off between the two kinds of IoT entities. It is more efficient and suitable for practical IoT applications.

\section{Conclusion}
Based on an improved CL-AS algorithm, we put forward an efficient verifiable data aggregation scheme for IoT scenarios, named VDAS, which effectively reduces the computation overhead in IoT data center. The scheme is proved secure against existential forgery under adaptively chosen messages attacks, and the security is tightly related to Computational Diffie-Hellman (CDH) problem in the random oracle model. Moreover, the size of aggregated authentication message is only two group elements, independent of the number of IoT terminals. The terminals perform individual signing operations in a non-interactive manner, using only their own secure information and public information of the system, which allows a legitimate terminal to dynamically participate in the verifiable data aggregate processing. Additionally, the proposed VDAS achieves a better trade-off on the computation overheads between the resource-constrained IoT terminals and the data center. The proposed VDAS is more suitable for the data authentication in the resource-constrained IoT environment such as the wireless sensor network, health-care system, VANETs data aggregation and so on.

\section*{Acknowledgements}
This work is supported by Natural Science Basic Research Plan in Shaanxi Province of China (No. 2016JM6057), National Science and Technology Major Project of the Ministry of Science and Technology of China (No. 2013ZX03005007), the 111 Project (B08038) of China, and the Qatar National Research Fund under grant NPRP 8-408-2-172.

\end{document}